\documentclass[10pt]{sensys-proc}

\usepackage{graphicx}
\usepackage{balance}
\usepackage{comment}
\usepackage{amsmath}
\usepackage{amssymb}

\DeclareMathOperator*{\argmax}{argmax}   % Jan Hlavacek

\numberofauthors{1}

\author{
\alignauthor Sam Royston \\
\affaddr{Department of Computer Science
}\\
\affaddr{New York University, 2016}\\
\email{sfoxroyston@gmail.com}
}

\title{Personalized Understanding of Blood Glucose Dynamics via Mobile Sensor Data}

\begin{document}

\maketitle
\begin{abstract}
Continuous Blood Glucose (CGM) monitors have revolutionized the ability of diabetics to manage their blood glucose, and paved the way for artificial pancreas systems. In this paper we augment CGM data with sensor input collected by a smart phone and use it to provide analytical tools for patients and clinicians. We collected GPS data, activity classifications, and blood glucose data with a custom iOS application over a 9 month period from a single free-living type-1 diabetic patient. This data set is novel in terms of it's size, the inclusion of GPS data, and the fact that it was collected non-intrusively from a free-living patient. We describe a method to measure the occurrence of lifestyle \textit{events} based on GPS and activity data, and show that they can capture instances of food consumption and are therefore correlated to changes in blood glucose. Finally, we incorporate these event representations into our system to create useful visualizations and notifications to aid patients in managing their diabetes.
\end{abstract}

% % A category with the (minimum) three required fields
% \category{H.4}{Information Systems Applications}{Miscellaneous}
% %A category including the fourth, optional field follows...
% \category{D.2.8}{Software Engineering}{Metrics}[complexity measures, performance measures]

% \terms{Mobile sensing, Health analytics}

% \keywords{ACM proceedings, \LaTeX, text tagging}

\section{Introduction}
\label{sec:intro}
Type-I Diabetes is an autoimmune disease characterized by the inability to produce insulin, and correspondingly, the inability to  regulate blood glucose. Diabetics must manually administer insulin to maintain stable blood glucose levels, and poor control has been linked to a host of dangerous side effects. Effective regulation of blood glucose is difficult and time consuming; our system offloads portions of the data-collection, regulation and prediction workload onto ubiquitous computing hardware. In addition, current trends in wearable tech point towards CGM technology becoming less invasive and possibly being eventually integrated into consumer fitness devices. Blood glucose is an informative variable that can help describe the way one's metabolism responds to a meal, or lack thereof; given a less-invasive CGM method, this information might prove useful to a variety of non-diabetics as well as the diabetics who depend on it.

We believe that as sensor "traces" generated by mobile devices become more ubiquitous, the following goals serve as important foundational principles for personal health analytics platforms aimed at diabetes and beyond. Correspondingly, the integrated system we are presenting is designed based on these tenets:

\begin{description}
    \item[Prioritizing effectiveness at the personal scale]{The sensitivity of health data may preclude large scale aggregate analyses in some cases, thus we should focus on approaches that work for $n=1$. Accordingly, we focus on a large dataset collected from a single, free-living type-1 diabetic. }
	\item[Continuous and unobtrusive data collection.]
{All data is collected \textit{autonomously} by a mobile phone and is then aggregated, cleaned, and analyzed on an remote server. Effects on the subject's life were minimal during data collection, and participation involved only turning on a mobile application such that it ran as a background process on their iPhone.}
    \item[Utilization of ubiquitous contextual data]{
        Data in the form of physical activity estimates, GPS location, and food purchase transactions are combined with data from a continuous blood glucose monitor (CGM) to provide representations and machine learning models that are interpretable and actionable.}\\
\end{description}

%%\subsection{The burden of \textit{Time Well Spent}}
Health application developers are beholden first and foremost to the needs and preferences of the patient.  For example, a state-of-the-art model using many external variables for blood glucose prediction is of little use to the patient if it relies on them to actively input a significant fraction of this data. In short, the patient's time is the ultimate currency and we hope to spend as little of it as possible. 

Let us define a \textit{Diabetes Care Action} (DCA) as a an action that a patient takes involving food intake, insulin dosage, or lack thereof. 
DCAs often involve checking one's blood glucose, if the tools are available.
To illustrate the trade-off between data collection and patient utility, consider a single DCA in which the patient inspects the CGM, insulin pump history, thinks about what they have eaten, and makes a decision for action, which could be either corrective or anticipatory.  
A typical DCA might be some combination of an insulin ``bolus'' dosage, corrective sugar intake, or pump basal rate update. 
A diabetic may make many DCAs throughout the course of the day, and depending on the quality of a DCA it may consume somewhere in the range of 15 seconds and 5 minutes of the patients time (a poor quality or "streamlined" action may be missing certain considerations, like checking the cgm or pump history). 
If we develop an analytics application that requires 20 minutes of data entry per day, we must be able to show that this time was better spent doing data entry instead of successfully completing diabetes care actions as described above. 
We anticipate that due to the patients exclusive knowledge of their own experiences and past metabolic reactions, this ``time well spent'' trade-off will be nearly impossible to ever justify in favor of data entry; therefore we aim to focus on data collection methods that take up virtually none of the patients time.  

The techniques required to make sense of the different types of data form a system designed to improve patient knowledge and diabetes management capabilities in real time. In this paper, our contributions are 
\begin{itemize}
	\item{Introduction of a large single-person data set which offers a novel combination of real-world predictors at a large scale}
    \item{A method for constructing discrete ``event'' features from unstructured smartphone sensor data which are correlated to changes in blood glucose\\}
\end{itemize}

\section{Motivation and Related work}
Previous work in the analysis and prediction of blood glucose data is comprised of many approaches. Machine learning research in this area is logistically challenging for a few reasons:
    \begin{description}
        \item[CGM benchmark Datasets are not publicly available] Due to the sensitive nature of medical data, datasets from CGM studies are often not released publicly. This creates a problem for reproducibility and prevents direct comparison of different algorithms. For example, a diabetic with good control will generate CGM data with less variance that may be easier to predict for a wide array of regression techniques. In this paper we use the error of static predictions (the last known value), is used to help characterize how ``easy'' the data is to predict.
        
        \item[CGMs are a black box] Continuous glucose monitors make use of proprietary techniques to estimate blood glucose levels via interstitial glucose levels. These systems likely include techniques like Kalman filters, but there is little explicit documentation from companies like Medtronic and Dexcom. 
        
    \end{description}

We can divide the prediction problem into two problem subtypes: 
\textbf{Hypoglycemia prediction} and \textbf{exact Blood Glucose prediction}. 
Hypoglycemia prediction frames the problem in a classification setting and
attempts to predict whether a window of measurements will result in 
hypoglycemia (low Blood Glucose) after within some time horizon. 
Predicting low Blood Glucose is of particular interest because it poses 
the most immediate threat to the well-being of the patient, and in 
especially bad cases, if left untreated can result in loss of 
conciousness, seizure, and rarely, death. Dassau et. al. uses a voting 
system combining a grab-bag of different predictors to achieve high 
accuracy in predicting hypoglycemic events 
\cite{dassau}.  

In contrast, exact Blood Glucose prediction is a regression 
problem and is often assessed via measuring the RMSE or $R^2$ error between the actual and predicted values. There have been approaches using artificial neural
networks \cite{perez}, as well as support vector machines and kalman filters \cite{marling}, 
\cite{marlingstudent}. 

In \cite{marling}, a metabolic model which incorporates 
insulin levels and food intake is fit to the data using a kalman filter, and 
the metabolic variables defined by this model are used as features for an SVM to make predictions. A prior attempt by [insert] to directly augment time-series blood glucose features with raw activity and dosage data for training a support vector machine for regression (SVR) was unsucessful and highlighted the importance of feature engineering in this problem. Marling et. al., built on this by using a metabolic model, along with patient recorded activities and insulin dosages to compose improved features for an SVM regressor. Marling et. al. reports a small but statistically significant gain between a version of their model which includes activity data, and one that does not. Rollins et. al. \cite{rollins} appoaches the problem from a signal processing perspective,
and data is recorded over 3 weeks from a type-2 diabetic wearing a bio-sensing body suit. 

A goal of many CGM predictive modeling studies is to aid in the development
of an artificial pancreas (AP). In pharmacological parlance, the \textit{kinetics}
of insulin absorption \cite{kinetics} and carbohydrate metabolism are slow enough such that 
classical control-theoretic frameworks are insufficient for effectively 
controlling blood glucose levels.  For this reason there has been interest in solving prediction tasks that could serve future 
artificial pancreas control systems.  \\

\section{Data Discovery and Prediction} 
    In this system we focus on the  of utilization of ``event'' features
constructed from location and activity data to improve our ability to predict blood glucose changes.

\subsection{Dataset}
We introduce a dataset which contains intermittent location, activity, and blood glucose measurements along with other meta-data collected from a 25 year old male type-1 diabetic. 
\begin{description}
	\item[Activity Classes (\textit{181,635} measurements)]{Activity classifications are performed continuously in the background even when the application is not running and thus are the most frequently collected data type. This is a categorical variable that is either \textit{'walking', 'running', 'cycling', 'stationary', 'automotive', 'stationaryautomotive',} or \textit{'unknown'}}
    
    \item[Locations (\textit{61,817} measurements)]{Locations were actively measured via GPS on the iPhone and sent directly to the server while the application runs in the background. The GPS polling policy is administered by the mobile operating system and generate more data when the device is in motion than when it is stationary, and no data at all if the device cannot make contact with the gps satellite}
    
     \item[Food Purchases]{The server queries various bank apis to catalog time-stamped food purchases, which may include other meta-data like the location of the purchase and it's cost}
    
    \item[Blood Glucose  (\textit{50,478} measurements)] Blood glucose levels (mg/dl) are measured every 5 minutes via a subcutaneous sensor by an external Dexcom CGM device, which communicates via bluetooth with the user's phone. \\
\end{description}

    Due to the wide range of sampling frequencies, we preprocessed the data by coercing it into an array of over 2.8 million feature vectors each representing a time-window of 15 seconds.  While the CGM test intervals are only every 5 minutes, by taking a snapshot every 15 seconds we were able to capture much more activity data, which is sometimes sampled at a rate upwards of 4 times per minute. 

\begin{figure}
    \begin{center}
    \includegraphics[width=1.1\columnwidth]{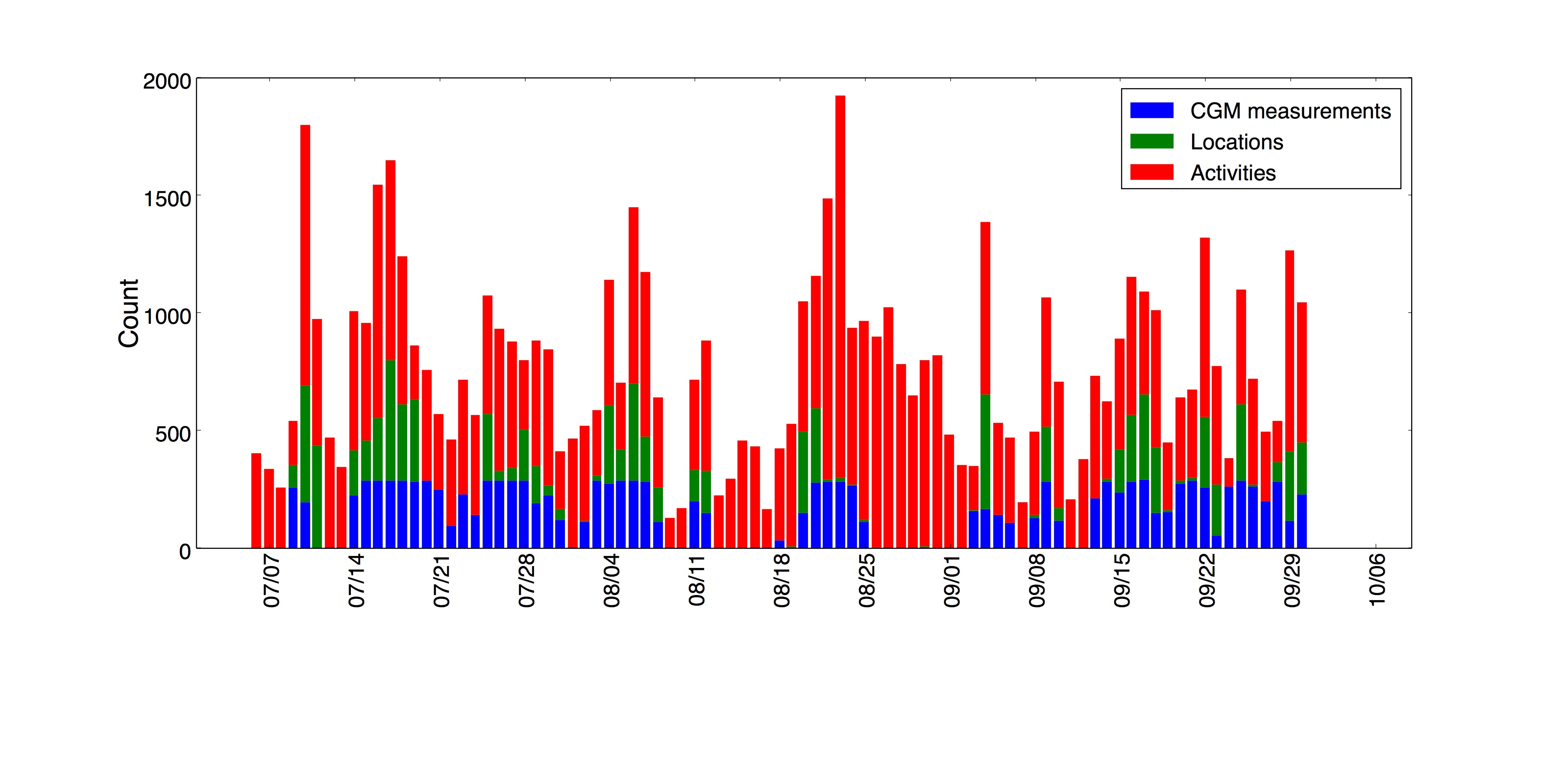}
    \end{center}
    \caption{Daily measurement frequencies}
\end{figure}

\subsection{Identifying Relevant Events}
An important first step to utilizing location and activity data is to identify candidates for when meals and other types of events occur.  Raw location and activity
data do not constitute viable features on their own, and does not contribute to prediction accuracy \cite{marling}, therefore we construct specialized features that are representative of visits to different locations.

\subsubsection{Activity Imputation}

As seen in Figure 1, while there is some regularity to CGM sampling, location and activity updates occur at irregular intervals determined by iOS internals which are dependent upon factors such as battery life and relative intensity of motion. Therefore robust imputation methods are needed for filling in the incomplete data. In this section we describe methods for imputing activity measurements. 

\begin{description}
    
\item[Nearest Neighbor]{One of the simplest and most effective methods for imputing activity data is by taking the nearest neighbor activity of the time slot we are trying to impute. However, this method is confounded by more complex patterns, e.g. activity measurements which often alternate. }

\item[Rolling Mode]{This method fills in all the missing activity predictions in a sample with the most frequent activity in that sample}\\
\end{description}

By using \textbf{Logistic Regression} for imputation, we
learn more complex relationships between activities by predicting the likelihood of each possible activity as a linear combination of activities over the sample time window. Concretely, our prediction of the activity at time $t$ is 
$$
\argmax_{a \in A} \big(\mathbf{x} \cdot \mathbf{w}_a \big)
$$
In addition, we can conveiniently compute the likelihood of a given activity $a$ as
$$\frac{\exp(\mathbf{x} \cdot \mathbf{w}_a)}{\sum_{i \in A} \exp(\mathbf{x}) \cdot \mathbf{w}_a}$$
Where $\mathbf{x}$ is a one-hot encoding of activities over some time window of width $2k +1$ centered at $t$,  and $\mathbf{w}_a$ is the $\vert A \vert \times 2k+1$ weight matrix corresponding to activity $a$. Thus the total set of weights $\mathbf{W}$ can be interpreted as a tensor of size $\vert A \vert \times \vert A \vert \times 2k+1$. 
The weights $\mathbf{W}$ are learned by minimizing the Negative Log-Likelihood loss defined in \cite{LeCun} via the back-propagation algorithm \cite{backprop}.
    An important aspect of this method is that we are left with scalar
likelihoods to which we can apply thresholding functions. 
\\ 
\begin{figure*}[t]
\begin{center}
\includegraphics[width=2.2\columnwidth]{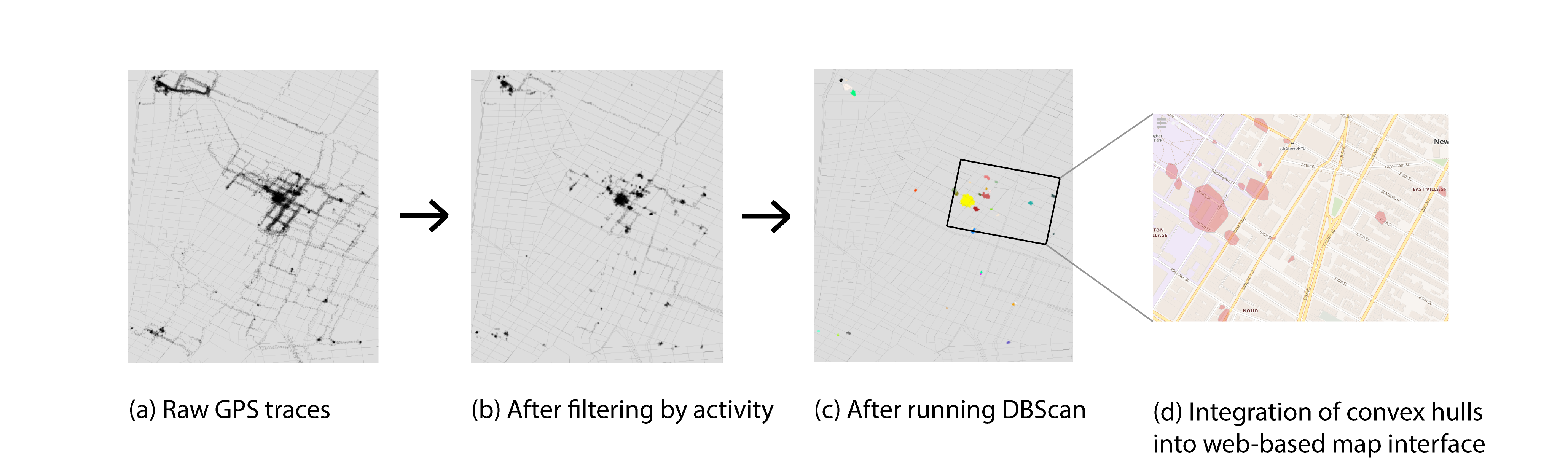}
\end{center}
\caption{Data flow diagram}
\end{figure*}

\subsubsection{Filtering and Clustering Locations}

By using one of the activity imputation methods described above we can filter most location measurements by their imputed activity at the time they were measured (although a few may be outside the range $k$ of any observed activities, in which case they cannot be imputed). In the case of the logistic regression imputation method we can filter locations by the likelihood associated with an activity at a given time. 

In this section we will be only filtering by \textit{stationary} location measurements, as these are likely to be taken when the user is consuming food. Granted, there are exceptions to this rule, but we treat these situations (i.e. eating while on the go) as unobserved events. 

To identify important events in a person's day, we must 
discretize the latitude/longitude space via clustering. \textit{K-Means}
clustering \cite{kmeans} is inappropriate in this scenario due to the fact that
in \textit{K-Means}, clusters must be convex and space filling. These
constraints are a problem since they do not conform to non-convex human
movement patterns and assign labels to areas with little or no data, which can
lead to poor generalizability. 
    Temporal methods specially designed for GPS data, such as those
described described in \cite{timeclusters} may also be useful for
clustering, but in practice we experinced difficulty in finding the correct
parameters for our setting which involved a multitude of visits to the same
locations.
    Density-Based clustering, or \textit{DBScan} is an agglomerative clustering
method with two parameters: $eps$ and $MinPts$ \cite{dbscan}.  DBScan builds
clusters $\mathbf{C}$ deterministically such that they satify the following
conditions for $\forall p,q \in C_i, \forall C_i \in \mathbf{C}$
\begin{enumerate}
    \item If $p \in C_i$ and there exists a path $p,q$ consisting of n points $p_i, ... p_n, p_i = p, p_n = q$ such that for $i < n$
            $
            \varsigma(p,q) = \vert N_{Eps}(p_i)\vert \geq MinPts \wedge p_{i+1} \in N_{Eps}(p_i)
            $
    is true, then $q \in C_i$
    \item  Noting that in general $\varsigma(p,q) \neq \varsigma(q,p)$, if $p,q \in C_i$ then $\exists o \in C_i$ such that $\varsigma(o,p) \wedge \varsigma(o,q)$ is true.\\
\end{enumerate}
    In words, the above conditions ensure that there is a path between all
points $p,q \in C_i$ which adheres to the density constraints parameterized by $Eps$ and $MinPts$. \textit{DBScan} Makes use of spatial hashing algorithms for efficiency and is ideal for our purposes since it discards points in areas of low density as ``noise'' and is capable of learning non-convex clusters.
 Although it should be noted that without first filtering by the inferred activity, identifying key locations via \textit{DBScan} becomes difficult due to spurious patterns (e.g. path intersections, where the user travels along
 paths frequently enough that their intersection is recognised as a cluster).
    
Once the clusters are determined by \textit{DBScan}, the next steps are to
further process the timestamps of the clustered locations as described in
section $3.2.3$, or to compute the convex hulls corresponding to each cluster 
for use in map interfaces and potentially for defining 
geo-fences.

Geometrically the clusters identified by this method correspond closely to actual restaurants and other frequented locations. However the points eventually used to define all of them are only a small subset of all the gps traces collected.
 \\
 \begin{figure}
 \begin{center}
\includegraphics[width=0.9\columnwidth]{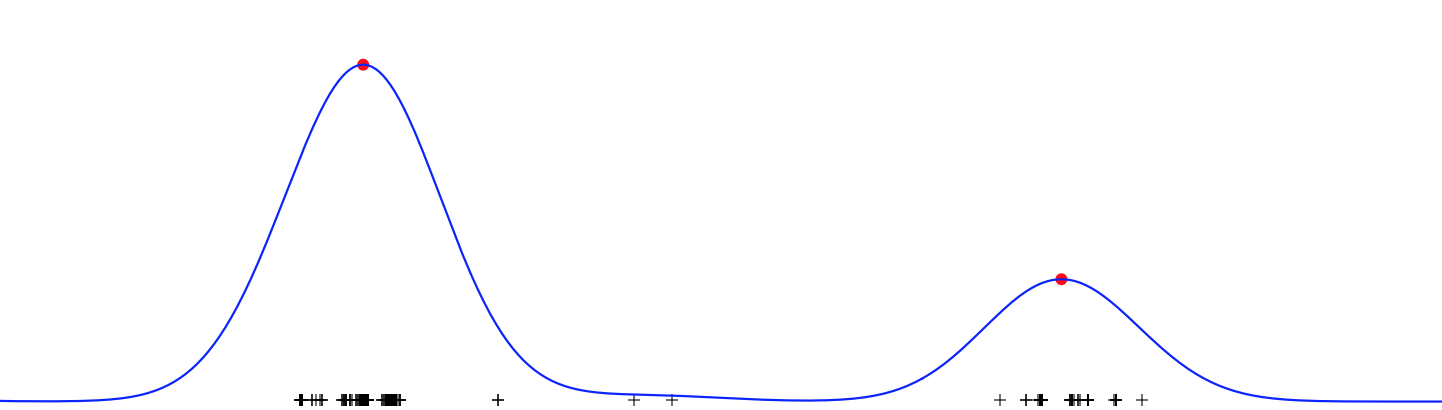}
\end{center}
\caption{Event discretization via kernel density estimation over a 3 hour period}
\end{figure}

 \subsubsection{Pinpointing When Events Occur}
    A specific ``hotspot'' identified by the algorithm is spatially defined by
a set of latitude-longitude pairs. However, it also corresponds to one or more discrete visits at this destination, each contributing a subset of the lat-lngs that constitute the cluster as a whole.
 The final step in identifying the key events in a user's day, is to associate a specific time with an occurance of the event: this task is analogous to one-dimensional clustering across time.
    First, to generate a smooth version of the observation impulses, we perform
Gaussian kernel density estimation, with a bandwidth $h$ of 2
hours. Gaussian Kernel Density Estimation computes a probability density
function $\rho(y)$ given measurement times $\mathbf{X}$ 
\begin{equation}
    \rho(y) = \sum_{x \in \mathbf{X}} \exp\Bigg( \frac{(y - x)^2}{h^2} \Bigg)
\end{equation}
The function $\rho(y)$ is smooth and it is trivial to find the set of local
maxima over a time interval. These maxima are used to align blood glucose
time-series resulting from different visits to the same location. In our
system, we align 2-hour windows of blood glucose readings, which allows us to
calculate aggregate blood glucose statistics associated with specific location
types.

There are multiple hyper-parameters within the event-identification framework, such as the $eps$ and $MinPts$ parameters of the $DBScan$ algorithm and the kernel width parameter of the density estimation step. These hyperparameters were selected to identify events that occur at the frequency of meals, rather than more frequent events like bathroom visits.  
\subsection{Understanding CGM Dynamics}
    In this section we consider the problem of analyzing the data
generated by the CGM device. We found that in the correct conditions, linear regression and its variants performed competitively, and outperformed other methods.

\subsubsection{Linear Models for Prediction}
    A simple linear regression model predicting a single variable
(blood glucose $t$ minutes into the future) selects the weights $w$
for a set of features so that we minimize the convex cost function
\begin{equation}
f(\mathbf{X}, \mathbf{y}) = \left\vert\left\vert \mathbf{w}^\top  \mathbf{X} - \mathbf{y}\right\vert\right\vert_2^2    
\end{equation}
Where $\mathbf{X}$ is the $n \times m$ matrix of $n$ samples with $m$ features and $\mathbf{y}$ is the vector of $n$ responses and $\mathbf{w}$ is the vector of $m$.
When we take our features to be the past $k$ blood glucose readings, the magnitude of the weights $\mathbf{w}$ can be interpreted as the importance of each past measurement in the estimate.
Variants of (1) such as the Lasso and Ridge Regression, also penalize the weights $\mathbf{w}$ by their $L_1$ and $L_2$ norms, respectively, in order to discourage the weights from learning random noise:\\
\begin{description}
    
\item[Lasso]{
\begin{equation}
f(\mathbf{X}, \mathbf{y}) = \left\vert\left\vert \mathbf{w}^\top  \mathbf{X} - \mathbf{y}\right\vert\right\vert_2^2 + \lambda \left\vert\left\vert \mathbf{w} \right\vert\right\vert_1
\end{equation}

The $L_1$ penalty promotes sparsity in $\mathbf{w}$. This is useful for preventing irrelevant weights from learning spurious noise (i.e. those for values far in the past).
}
\item[Ridge]
\begin{equation}
f(\mathbf{X}, \mathbf{y}) = \left\vert\left\vert \mathbf{w}^\top  \mathbf{X} - \mathbf{y}\right\vert\right\vert_2^2 + \lambda \left\vert\left\vert \mathbf{w} \right\vert\right\vert_2^2
\end{equation}

The $L_2$ penalty constrains the size of the wieghts $\mathbf{w}$.\\

\item[Elastic Net]
\begin{equation}
f(\mathbf{X}, \mathbf{y}) = \left\vert\left\vert \mathbf{w}^\top  \mathbf{X} - \mathbf{y}\right\vert\right\vert_2^2 + \alpha \big ( \left\vert\left\vert \mathbf{w} \right\vert\right\vert_2 + \lambda \left\vert\left\vert \mathbf{w} \right\vert\right\vert_1 \big )
\end{equation}

The elastic net uses a weighted combination of the losses employed in Lasso and Ridge regression.\\

\item[Total Variation Regularization]
\begin{equation}
f(\mathbf{X}, \mathbf{y}) = \left\vert\left\vert \mathbf{w}^\top  \mathbf{X} - \mathbf{y}\right\vert\right\vert_2^2 +  \lambda \left\vert\left\vert \triangledown(\mathbf{w}) \right\vert\right\vert_1 \big )
\end{equation}

TV regularization induces sparsity on the finite-differences operator of the weights $\triangledown(\mathbf{w})$ .  Intuitively, variation regularizing priors make sense for time series applications like this one, because we can expect that measurements are increasingly similar if taken within a shorter time-span.\\
\end{description}
    The models above come with advantages and disadvantages. Each of them defines an efficiently-solved convex optimization problem on $\mathbf{w}$, but a weakness is that they assume a fixed number of observations per sample. \\
\subsubsection{Non-parametric Kernel Regression}
In order to work with unevenly sampled data, we use the Nadaraya and Watson \cite{watson} formulation of \textit{kernel regression} as described in \cite{narges}.
We would like to compute the expected value of a missing variable at any point in time, with no assumptions about the spacing of the observations.

\begin{equation}
    f(\mathbf{X}, \mathbf{y}) = \sum \Bigg( \frac{\mathbf{w}^\top  \mathbf{x}_i}{\mathbf{w}^\top  \mathbf{\delta}_i} - y_i\Bigg)^2  
\end{equation}

Where $\mathbf{\delta}_{i}$ is the same size as a sample vector $\mathbf{x}_i$ and for each $d_{j} \in \mathbf{\delta}_i$ and $x_j \in \mathbf{\delta}_i$  
$$
 d_{j} = \begin{cases} 
                                1 & x_{j} \text{ contains an observation} \\
                                0 &  \text{otherwise}
       \end{cases}
$$

In short, $\mathbf{\delta}_{i}$ simply marks which time-slots contain a measurement, regardless of what that measurement it. We can imagine the denominator in (7) as a normalization term which allows for effective learning from data with a variable number of observations. By normalizing with $\mathbf{w}^\top \cdot \mathbf{\delta_{x}}$ we ensure the estimate of $\mathbf{y}$ does not scale with the number of observations. Note that when $\mathbf{w}$ is orthogonal to a vector c $\delta_{obs}$ the loss function (6) blows up, thus implying that (6) is not convex. The situations for which $\mathbf{w}^\top \cdot \mathbf{\delta_{x}} = 0$ can happen in as many ways as there are unique $\mathbf{\delta}_i$'s. Therefore, learning $\mathbf{w}$ via stochastic gradient descent and back-propagation is a natural choice, as these are techniques often used in non-convex settings. Back-propagation has seen recent success applied to training deep neural networks for image classification, but can be applied to learn weights using any differentiable objective. 
Furthermore, this approach is advantageous in that SGD is an on-line algorithm and therefore does not assume a stationary signal. In fact, we could learn the weights of the linear regression models in section 3.3.1 in an online fashion by using back-propagation, however convergence time might be slower. The main  disadvantaage of this approach is that it does not enjoy the guarantees of the convex cost functions discussed in \textit{3.3.1}.

\subsubsection{Incorporating Exogenous features}
    The filtered labels which result from the techniques in section 3.2 can
be transformed into a one-hot encoding which
describes when then user is stationary at an oft visited location. It remains to
determine which locations are significant for blood glucose prediction. In this study, we simply picked a subset of clusters by hand which are food consumption locations as ``significant'' (i.e. deli, thai food, pizza), and ignored the rest. This approach has the issue that every observed value is equal to 1 (i.e. True) so that that normalization term (in 6) for the onehot-encoding is the same as numerator. To recast the one-hot encoding as a vector binary valued observations we count any location measurement or purchase as an observation, even in the event that the location is not deemed a ``relevant event''.       
An interesting area for further research is the relationship between the proportion of missing values and the performance gain of (5) over (1).

\begin{equation}
    f(\mathbf{X}, \mathbf{y}) = \sum \Bigg( \frac{\mathbf{w_{bg}}^\top  \mathbf{x_{bg}}_i + \mathbf{w_{exog}}^\top  \mathbf{x_{exog}}_i}{\mathbf{w_{bg}}^\top  \mathbf{\delta_{bg}}_i +  \mathbf{w_{exog}}^\top  \mathbf{\delta_{exog}}_i} - y_i\Bigg)^2  
    \Gamma
\end{equation}

\section{System}
The system is comprised of a web server, mobile application, and an external CGM device manufactured by Dexcom, and handles data collection, analysis, and reporting information to the user. In order to allow more freedom in selecting machine learning tools, all of the analysis workload is performed remotely rather than on the user's device. 
\begin{figure}[t]
\begin{center}
\includegraphics[width=1.03\columnwidth]{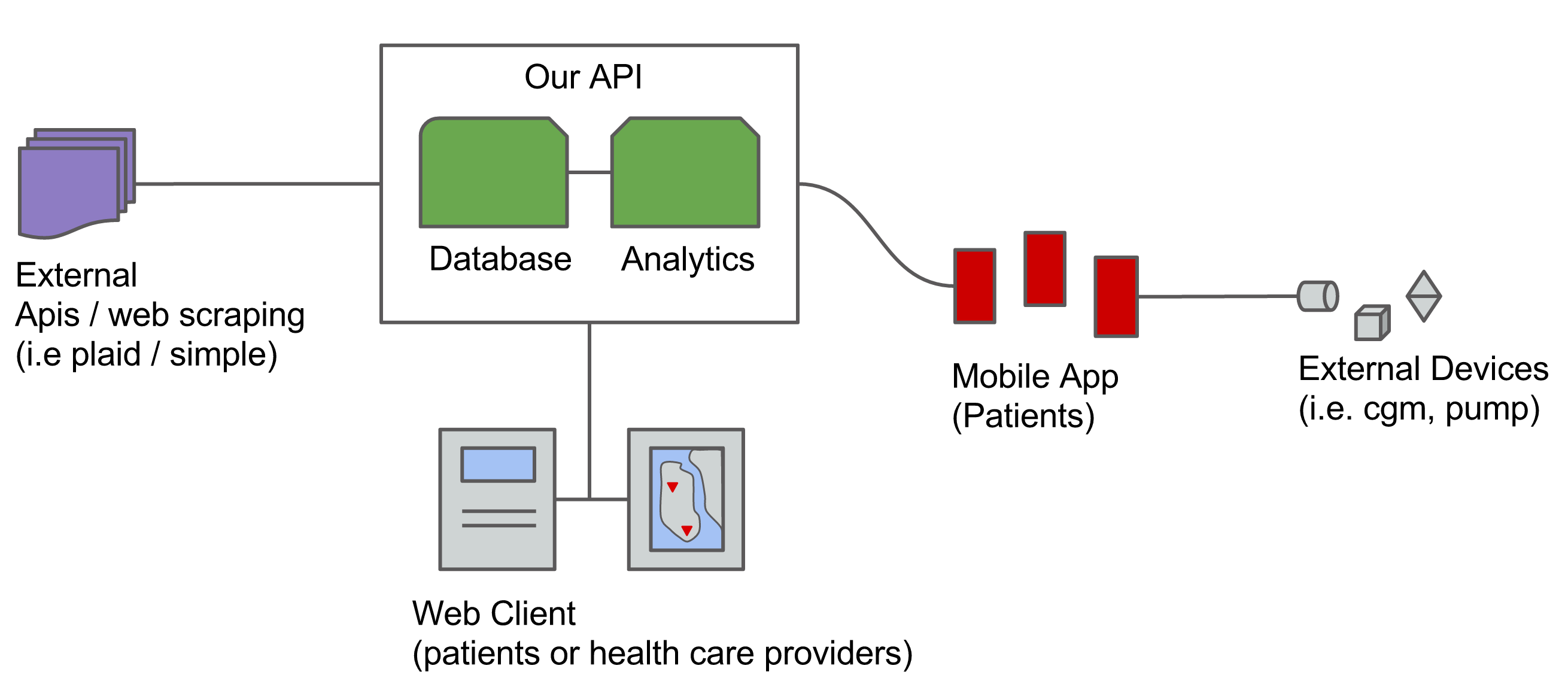}
\end{center}
\caption{Data flow diagram}
\end{figure}

A significant priority is that our system is able to provide real time analysis and predictions to the user but there are a few hurdles to achieving this. 
\begin{itemize}
    \item{ Some data, like credit card purchases, is not updated in real-time }
    \item{ There are constraints imposed by the device maker on background app usage, due to battery usage and privacy considerations \\ }
\end{itemize}

    All of the software currently is deployed in one of two places: the mobile application,
or the web-server back-end.

\subsection{ Mobile Application }
The majority of the relevant data collection is done via the mobile application, which runs on an iphone and was written in Objective-C. 
The mobile application interfaces with Apple's HealthKit framework to access data recorded by the CGM and transmitted to the phone via bluetooth. The CoreLocation framework is used to receive frequent high-accuracy location updates as well as set geo-fences for passive tracking.  

    Interestingly, Apple's hardware accelerated activity
classification system automatically stores and records inferred activities ranging back 7 days in a local database on the device. This means the while to record location data the application needs to be running in the background, activity data is recorded independently. This feature partly explains the frequency of activity measurements relative to the other data sources.  Upon startup, our app synchronises and uploads the activity data that was collected since it last ran.

\subsection{Location Tracking}
    By far the biggest issue with the mobile data-collection system is 
the effect of constant location tracking on the battery life of the 
phone. Although Apple has made considerable optimizations to the 
collection of location data, which is one reason it is not sampled 
uniformly, battery life does not exceed 24 hours when tracking is 
enabled, and in some cases can be less. The rate of collection of GPS coordinates is determined by 
the user's speed, likely because stationary location updates are not 
useful for most applications. A mobile user will cause location 
updates on the order of 1 per 10 - 15 seconds. If the application 
adversely effects the battery life of the users device, the utility 
of the overall system goes down, therefore we employ a geo-fencing 
strategy that is more power efficient. However this means that the mobile application assumes two distinct data-collection modes: 
\begin{description}
    \item[Active Tracking]{where raw GPS traces are collected to form new clusters or ``hotspots'}
    \item[Passive Tracking]{where the application only reacts to entrances to and exits from previously learned hotspots.\\}
\end{description} 
    Implementing passive tracking is relatively straightforward in
iOS, as the CoreLocation framework provides a geo-fencing api. Apple restricts the number of active fences per application to 20, which was not enough to track all the learned clusters in our experience. To work around this we use a nested geo-fence strategy implemented locally on the device to actively update the tracked regions based on the larger \textit{superfence} the user is currently situated in. 

\subsection{Server}
    Data such as debit card food purchases, which were used in some
exploratory analyses, were gathered recurrently by the server.
Gathering data from different and sources means that significant work must be done to format the data for ingestion into the machine learning components, long-term storage,
and into the correct format for quick downloads to populate front-end visualizations.
    The server application needs to balance the versatility of a
structured database, while storing processed copies of data to be quickly sent to client applications.
Each individual measurement is indexed and stored in a MongoDB instance so that complex queries can be performed when needed. However, for most data requests from the user to the server, responses consist of preprocessed, compressed files stripped of irrelevant metadata.

\subsection{Data Processing}
Much of the data we are concerned with is not reported at uniform intervals and from separate sources; this means matching timestamps is sometimes the only means to match cooenciding measurements. Relying on timestamps can introduce problems; like if the system clock of a single component gets out of synch with the rest. This was a problem in our system on a few occasions. A possible approach for reducing the number of such failures is to mandate that a single (software) component drives communication with the server, as opposed to the remote database being updated individually by multiple gro on the device. This way data will arrive at the database already grouped with cooenciding entries, and the data cross-referencing point of failure will be  at least parially removed.

\subsection{ Interface }
    There are two user interfaces in the application: web and mobile. The mobile
application primarily acts as a data collection platform rather than a data visualization toolkit, but nonetheless includes a map interface. 
The web application frontend is the primary interface for data visualization and takes advantage of MapBox's webgl support for map annotation. Using these tools it is possible to draw 50,000+ lines on a tiled map interface without a significant performance slowdown. By displaying each and every portion of the user's recorded path, we aim to make the experience fun and provide a means to jog the user's memory.  We utilize browser based filtering to determine the range over which to display gps information. 

\begin{figure}[t]
\begin{center}
\includegraphics[width=1\columnwidth]{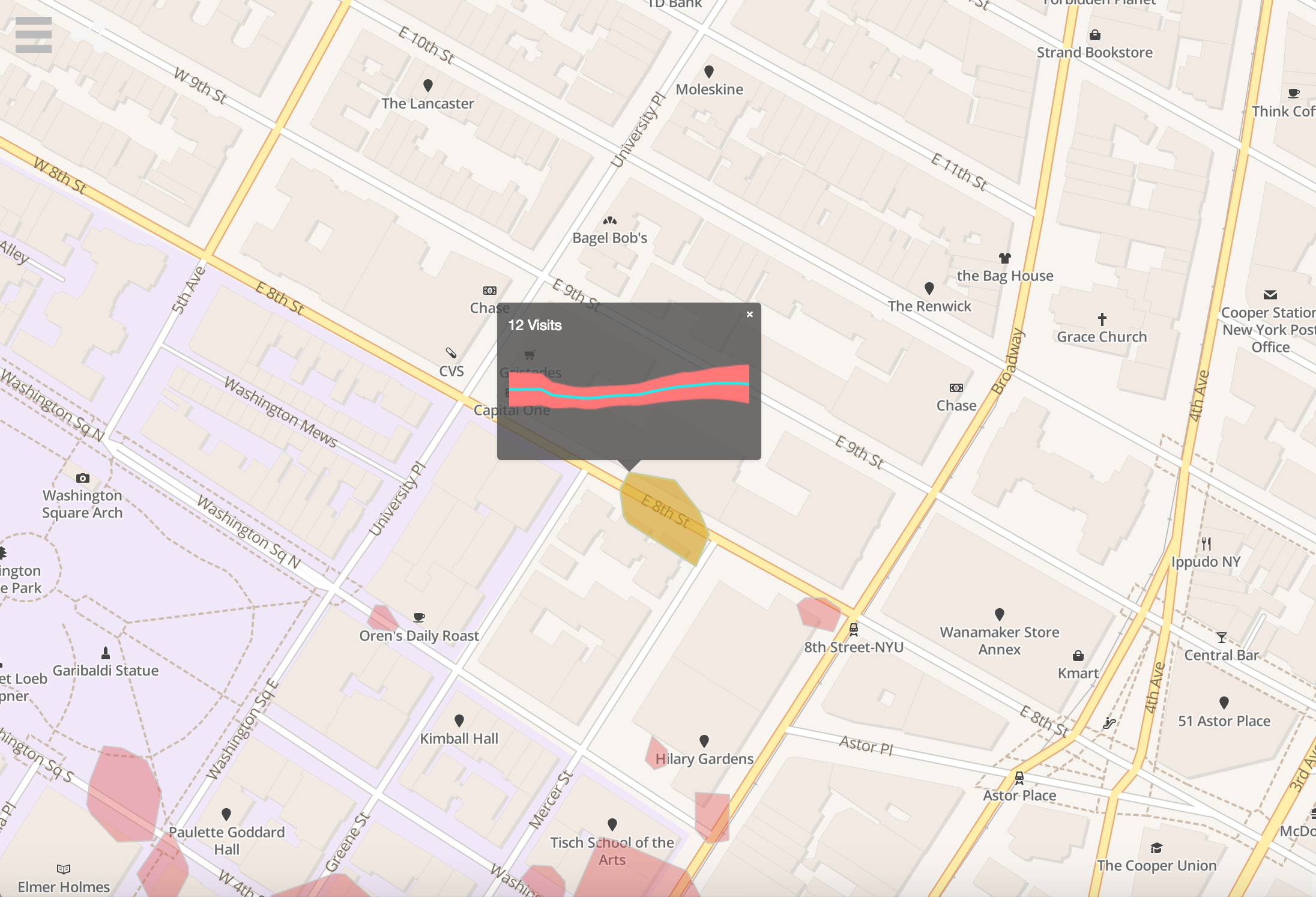}
\end{center}
\caption{Map Interface showing average and standard deviation of blood glucose during and after visits at an identified ``hotspot''}
\end{figure}

The relevant clusters determined via the method described in section 3.2 are drawn as convex hulls on the map interface and respond to user interactions. By selecting a cluster, or ``hotspot'' the user can view aggregate CGM statistics corresponding to all the visits at that location.

\begin{figure*}[t]
    \begin{center}
        \begin{tabular}{||c|c|c|c|c||}
        \hline
        Activity & Sample Mode & Nearest Neighbor & Logistic Classifier &   \\ \hline\hline
        Overall Accuracy     &  0.78 & 0.79 & \textbf{0.87} & ~  \\ \hline\hline
        Stationary           &  0.89 & \textbf{0.96} & 0.95 & ~  \\ \hline
        Automotive           & 0.23  & 0.07 & \textbf{0.83} & ~  \\ \hline
        Stationary Automotive& 0.12  & 0.02  & \textbf{0.46} & ~\\ \hline
        Walking              & 0.15  & \textbf{0.44}  & 0.35 & ~ \\ \hline
        Cycling              & 0.13  & \textbf{0.53}  & 0.34 & ~          \\ \hline
        \end{tabular}
    \end{center}
    \caption{F-Scores of each imputer for each activity category (excluding running, because the number of cases in the test set was so small)}
\end{figure*}

\section{Results}
We used the dataset described in 3.1 to evaluate a battery of prediction 
methods and interpret the models that are learned. For comparing the 
different prediction methods, we restricted the test sets to contain only 
complete observations so that all methods could be compared. In the case of 
kernel regression, the training set contained incomplete samples as well as 
the complete ones that the other methods were trained on. We show that 
kernel regression is able to incorprate the incomplete data to improve 
prediction performance.\\

\subsection{Activity Prediction}
We used a logistic classifier to impute activities using a window of recent activities as input. We compared the per class f-score (1-vs-all) to the scores of rolling mean and nearest neighbor baselines. We also computed the overall accuracy of each type of imputer.
The Logistic classifier outperformed the two baselines overall, largely because it's ability to learn about the frequent alternation of the Automotive and Stationary-Automotive measurements. For predicting more rare measurements like Walking and Cycling, nearest neighbor performed the best. The main advantage of the logistic classifier system is it's ability to produce likelyhoods of a given activity for a given time, which comes in handy during the filtering step described in section $3.3.2$. \\

\begin{figure}[t]
\begin{center}
\includegraphics[width=1\columnwidth]{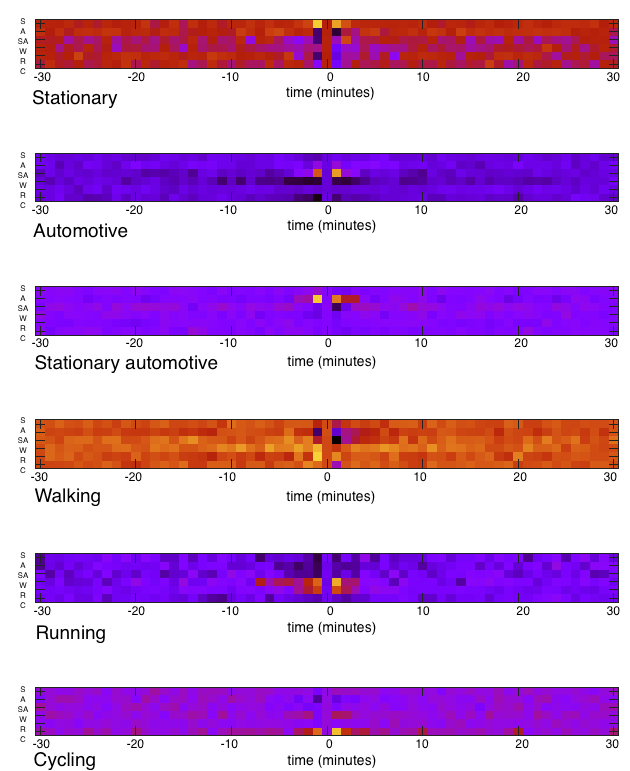}
\end{center}
\caption{The weights $\mathbf{W}$ which are used to impute activities. Note that Automotive signals overweight Stationary-Automotive signals as a predictor and vice versa. The is due to the fact that they often alternate.}
\end{figure}

\subsection{Event Identification}
In section $3.2$ we outlined a method for identifying discrete events within a persons routine using ubiquitous sensor data. In order to show that these events are relevant to diabetes care and analysis, we computed the Pearson correlation between a given type of lifestyle event occurring and the event of blood glucose increasing more than $30 mg/dl$ over a given time interval. We compared the correlations over different time intervals and types of events. Unsurprisingly, the group of events with the highest correlation was the one selected by manually referencing cluster boundaries against locations which were known to be associated with food consumption.  While it is not completely clear why general lifestyle event occurrences are more correlated than instances of ``stationary'' behavior, as detected by the accelerometer, but possible explainations include 

\begin{description}
    \item[A bias toward recurrent events] It could be the case that the subject is more likely to eat or ingest sugar at a place they have already visited, than at a new location. The lifestyle events described in section 3.2 filters by density and therefore is less likely to identify when the user visits a new location for the first time. If it is true that the user is a creature of habit, then the methods described are more appropriate for detecting when they ingest food.
    \item[Application activation bias] Lifestyle events can only be measured if both activity estimates and location data are available. While activity estimates are made regardless of whether the application is running, the application must be running in the background for location to be tracked. The user may have been biased towards ensuring that the application was active during food ingestion, simply because this is probably the most important event we were trying to track in the study, thus leading to a slight correlation between the presence of location measurements and an increase in blood glucose.\\
\end{description}

In addition, in figure 8 we can see that the maximum correlation is around 60 minutes after the detected event. This makes sense due to to fact that it takes at least 10-15 minutes for any blood glucose increase from food ingestion to be detected by the CGM, and some types of food may lead to gradual increases for hours afterwards.

\begin{figure}[h]
\begin{center}
\includegraphics[width=1\columnwidth]{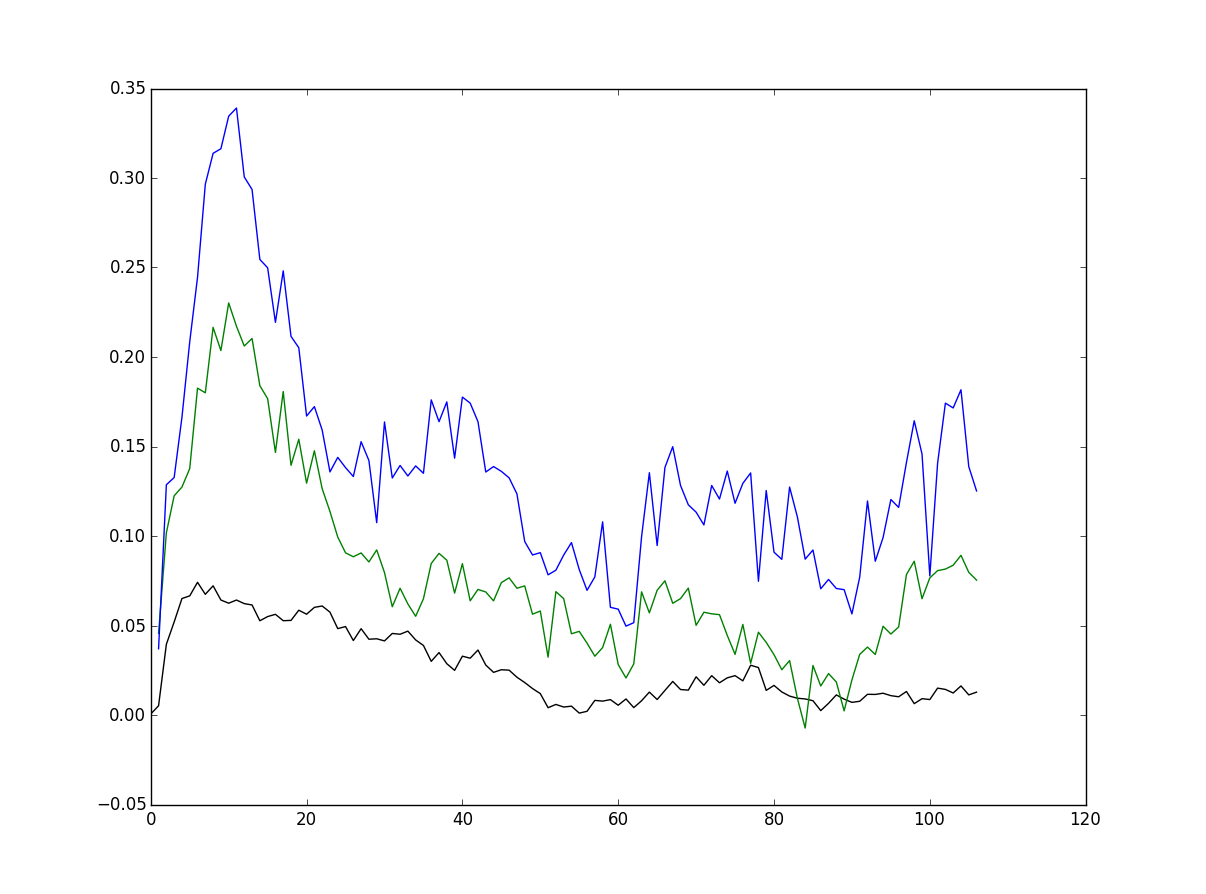}
\end{center}
\caption{Correlation with blood glucose over time following different non-BG signals: ``stationary'' measurements (black), lifestyle event occurrence (blue), and lifestyle event occurrence filtered by whether the cluster corresponds to a food consumption location (green).}
\end{figure}
\subsection{Blood Glucose Prediction}
    The data was pre-processed so that it had 
zero-mean and unit-variance. In our experience, linear regression models were surprisingly robust in their predictions and outperformed Support Vector machines both when lagged values are used as 
features and when features like slope and curvature are used for the SVM. Furthermore, among the linear models total-variation regularization had a slight edge. While Nonparametric kernel regression performed well, there was no difference in the performance between the model using only past blood glucose data (7) and the model which used exogenous information (8). In general, using exogenous data sources in prediction tasks proved to be difficult, and often led to lower performance. Therefore, we would reserve the usage of the features described in section 3.2 for assisting the user in their understanding the relationships between their lifestyle, food habits, and blood glucose dynamics. Figure 8 demonstrates that there is indeed a significant relationship, especially among hotspots designated at possible food locations.

\begin{figure}[h!]
    \begin{center}
        \begin{tabular}{||c|c|c|c|c||}
        \hline
        Horizon & Static & Linear  & Lasso & Ridge\\
          ~& ~ &  Regression & $\lambda = 1.0$ & \\
          \hline\hline
        15      &  14.3  & 10.42             & 10.43              & 10.42                \\ \hline
        30      & 25.86  & 20.5              & 20.49              & 20.49              \\ \hline
        60      & 43.52  & 37.64             & 37.63              & 37.64                \\ \hline
        \end{tabular}
    \end{center}
\end{figure}

\begin{figure}[h!]
    \begin{center}
        \begin{tabular}{||c|c|c||}
        \hline
        Elastic Net & TV  & Nonparametric Kernel  \\
        $\alpha = 1.0$, $\lambda = 0.5$ &  Regularization & Regression\\
        \hline\hline
        10.43    & -      &\textbf{10.42}             \\ \hline
        20.49    & 20.4   & \textbf{18.04}                 \\ \hline
        37.64    & -      & \textbf{31.89}               \\ \hline
        \end{tabular}
    \end{center}
    \caption{Cross validation results for blood glucose prediction}
\end{figure}

\section{Conclusions}
We have described a system for collecting contextual data along with blood glucose measurements, and a method for generating useful features from that data. Furthermore, we have described benchmark blood glucose prediction results using linear models at different time horizons. The main purpose of this work is to demonstrate the possibilities enabled by collecting contextual data along with blood glucose data. The ``event'' features allow for concrete association of blood glucose statistics with frequently visited locations. We hope that given more data, contextual features like those described in this paper may contribute to prediction and control systems.  \\
%ACKNOWLEDGMENTS are optional


\bibliographystyle{abbrv}

\bibliography{sigproc} 

\begin{thebibliography}{10}

\bibitem{marling}
R.~Bunescu, N.~Struble, and C.~Marling.
\newblock Blood glucose level prediction using physiological models and support
  vector regression.
\newblock {\em International Conference on Machine Learning and Applications},
  2013.

\bibitem{perez}
P.-G. C, F.~A, S.~G, C.~C, G.~EJ, R.~M, de~Leiva~A, and H.~ME.
\newblock Artificial neural network algorithm for online glucose prediction
  from continuous glucose monitoring.
\newblock {\em Diabetes Technol Ther.}, 2010.

\bibitem{dassau}
D.~E, C.~F, L.~H, B.~BW, Z.~H, J.~L, C.~HP, W.~DM, B.~BA, and D.~F. 3rd.
\newblock Real-time hypoglycemia prediction suite using continuous glucose
  monitoring: a safety net for the artificial pancreas.
\newblock {\em Diabetes Care}, 2010.

\bibitem{rollins}
M.~Eren-Oruklu, A.~Cinara, D.~K. Rollins, and L.~Quinn.
\newblock Adaptive system identification for estimating future glucose
  concentrations and hypoglycemia alarms.
\newblock {\em Automatica}, 2012.

\bibitem{dbscan}
M.~Ester, H.-P. Kriegel, J.~Sander, and X.~Xu.
\newblock A density-based algorithm for discovering clusters in large spatial
  databases with noise.
\newblock {\em KDD}, 1996.

\bibitem{timeclusters}
J.~H. Kang, W.~Welbourne, B.~Stewart, and G.~Borriello.
\newblock Extracting places from traces of locations.
\newblock {\em Proceedings of the 2nd ACM international workshop on Wireless
  mobile applications and services on WLAN hotspots}.

\bibitem{kmeans}
T.~Kanungo, D.~M. Mount, N.~S. Netanyahu, C.~D. Piatko, R.~Silverman, and A.~Y.
  Wu.
\newblock An efficient k-means clustering algorithm: Analysis and
  implementation.
\newblock {\em IEEE Transactions on Pattern Analysis AND Machine Intelligence},
  2002.

\bibitem{backprop}
Y.~LeCun, L.~Bottou, G.~B. Orr, and K.-R. Muller.
\newblock Efficient backprop.
\newblock Technical report, Image-Proccessing Research Dept. AT\&T Labs.

\bibitem{LeCun}
Y.~LeCun, S.~Chopra, R.~Hadsell, M.~Ranzato, and F.~J. Huang.
\newblock A tutorial on energy-based learning.
\newblock {\em to appear in Predicting Structured Data}, 2006.

\bibitem{narges}
N.~Razavian and D.~Sontag.
\newblock Temporal convolutional neural networks for diagnoses from lab tests.
\newblock {\em Working Paper}, 2015.

\bibitem{kinetics}
R.~S. Sherwin, K.~J. Kramer, J.~D. Tobin, P.~A. Insel, J.~E. Liljenquist,
  M.~Berman, and R.~Andres.
\newblock A model of the kinetics of insulin in man.
\newblock {\em The Journal of Clinical Investigation}, 1974.

\bibitem{marlingstudent}
N.~W. Struble.
\newblock Measuring glycemic variability and predicting blood glucose levels
  using machine learning regression models.
\newblock Master's thesis, Ohio University, 2013.

\bibitem{watson}
G.~S. Watson.
\newblock Smooth regression analysis.
\newblock {\em The Indian Journal of Statistics}, 1964.

\end{thebibliography}
% sigproc.bib is the name of the Bibliography in this case
\end{document}